\documentclass[doublecol]{epl2} 
\usepackage{amsmath,amssymb}
\usepackage{graphicx}

\title{Effect of disorder on a Pomeranchuk instability}
\shorttitle{} %Insert here a short version of the title if it exceeds 70 characters

\author{A.~F.~ Ho\inst{1} \and A.~J.~Schofield\inst{2}}
\shortauthor{A.~F.~Ho  \etal}

\institute{                    
  \inst{1} Department of Physics, Royal Holloway, University of London, Egham, Surrey TW20 0EX, UK.\\
  \inst{2} School of Physics and Astronomy, University of Birmingham, Birmingham B15 2TT, UK.}

\abstract{
We study the effect of weak and dilute disorder on the order parameter equation and
transition temperature of a Pomeranchuk-type Fermi-surface instability
using replica mean field theory. We consider the example of a phase
transition to a {$d_{x^2 -y^2}$} type Fermi surface distortion, and show
that, in the regime where such a transition is second order, the
transition temperature is reduced by disorder in essentially the same
way as that for a d-wave superconductor. We argue that observing this
disorder dependence of metal-to-metal transition is a useful
indicator of a finite angular momentum Fermi surface distortion.}

\begin{document}

\maketitle

\section{Introduction}

In recent years a growing number of correlated systems have been found
to exhibit thermodynamic phase transitions between metallic
states. Examples include the 17K transition in $\rm
URu_2Si_2$~\cite{palstra_1985a} and the transitions around the
metamagnetic quantum critical endpoint of $\rm
Sr_3Ru_2O_7$~\cite{grigera_2004b}. In contrast to the more familiar
superconducting or magnetic instabilities, the order parameter which
presumably develops at these transitions appears to be transparent or
only weakly coupled to most experimental probes.  Thus identifying the
nature of this ``dark'' or ``hidden'' order is a
challenging problem. Many years ago
Pomeranchuk~\cite{pomeranchuk_1958a} found a condition for
instabilities between metallic states characterized by Fermi surface
shape distortions. It has been argued that this instability occurs in
quantum Hall systems~\cite{fradkin_1999a,fradkin_2000a} and is the
origin of the transitions  in ${\rm URu_2Si_2}$~\cite{varma_2006a}
and ${\rm Sr_3Ru_2O_7}$~\cite{grigera_2004b}. However, the key
question remains how to identify  this sort of
order--particularly if bulk changes are masked by domains formation. In this Letter we calculate the form of the disorder dependence
of the transition temperature and find it to have a characteristic signature
of momentum space distortions of the metallic
Fermi surface.

We are motivated by the empirical similarity in the way the mysterious
phase in the bilayer ruthenate, ${\rm
Sr_3Ru_2O_7}$~\cite{grigera_2004b}, and the superconductivity in the
related single-layer compound, ${\rm
Sr_2RuO_4}$~\cite{mackenzie_1998b}, are both eliminated with very low
levels of disorder. In the latter case quantitative comparison of the
strong disorder dependence of the superconducting transition
temperature ${T_c}$ to the well-known form~\cite{larkin_1965a} has
become the {\em de-facto} signature of non-zero orbital angular
momentum pairing (in the absence of phase sensitive methods).  In
contrast $s$-wave superconductors are insensitive to
disorder~\cite{anderson_1959a}. Could a similar dependence be used to
diagnose the Pomeranchuk transition?

In this Letter we show how the Pomeranchuk instability is an analogue
of non $s$-wave superconductivity but in the particle-hole rather than
particle-particle channel. We exploit this using a combination of
standard methods to show that, not only might one similarly expect a
sensitivity to disorder but that the precise form of the disorder
dependence of $T_c$ is, under certain circumstances, {\em identical}
to that of unconventional superconductors.  This provides a
quantitative test of the ``dark order'' metallic phase which parallels
that now used for unconventional superconductivity.  Moreover, since
we show that the Pomeranchuk instability is strongly suppressed in
presence of weak disorder, our results may provide an explanation as
to why this rather subtle metal to metal transition is not observed
more often in nature.

\section{Model}

To study effects of disorder in a simple model of a Pomeranchuk
instability, we consider electrons on a two dimensional (2D)
tight-binding lattice with  a quadrupolar interaction that
has been studied extensively~\cite{oganesyan_2001b, kee_2003b,
khavkine_2004a, doh_2006a}
\begin{eqnarray} 
\label{Hpomm}
H_{\rm int} = \sum_{\bf k, p, q} \frac{1}{2} 
V_{\bf q}({\bf k}, {\bf p}) \; \psi^{\dagger}({\bf k+q}) \psi({\bf k}) \;
\psi^{\dagger}({\bf p-q}) \psi({\bf p}) .
\end{eqnarray} 
Here $\psi^{\dagger}_{\bf k} = [\psi^{\dagger}_{\bf k \uparrow} \;
\psi^{\dagger}_{\bf k \downarrow}]$ is the spinor creation operator.
The interaction explicitly has angular momentum dependence: $ V_{\bf
q}({\bf k}, {\bf p}) = g_{\bf q} \phi_{\bf k} \phi_{\bf p}$, where the
$d_{x^2 - y^2}$ form factor $\phi_{\bf k}= \cos k_x - \cos k_y$. 
Kee {\it et al.}~\cite{kee_2003b} found that
the $d_{xy}$ component of the quadrupolar interaction usually does not acquire an expectation value, which we thus drop.
%This separable potential allows a simple mean-field decoupling.

We add a weak, dilute  disorder potential that couples to the electron density
$H_{\rm dis} = \int d^d x \xi(x) \;\psi^{\dagger}(x) \psi(x)$
assuming, for simplicity, delta-correlated, static (quenched) disorder
${\rm Pr}(\xi) \propto \exp -\int d^d x \;\xi^2(x)/2 D $. Here $D =
1/2\pi N_0 \tau$ is the strength of the disorder potential, with $N_0$
the density of state at the Fermi surface, and $\tau$ is the disorder
scattering time.  We do not here consider stronger or correlated disorder,
where the disorder potential
may couple directly to the Fermi surface distortion (Pomeranchuk) 
order parameter\cite{carlson_2006a}.

\section{Methods}

For quenched disorder, one needs to disorder-average the free energy
instead of the partition function. One standard method that works also
for interacting systems is the replica
trick\cite{edwards_1975a,belitz_1994a}, based on the identity: $\ln Z
= \lim_{n\rightarrow 0} (Z^n -1)/n$ .  Note that we have also derived
the results presented below using diagrammatic perturbation
theory~\cite{abrikosov_1963a} but we found that the replica method
makes the parallel with unconventional superconductivity explicit.
The idea is to replicate $n$ copies of the partition function $Z$,
disorder-average $Z^n$, and finally take the limit $n\rightarrow 0$ to
get the disorder-averaged free energy. Since we have taken a simple
Gaussian distribution for the disorder potential, this
disorder-average is readily done to give the disorder-induced
interaction 4-fermion term (in momentum representation)
\begin{eqnarray} 
S_{dis} &= & \frac{-1}{4\pi N_0 \tau}
\sum_{\alpha,\beta}{T^2} \sum_{n, m}\;\; \sum_{\bf k_1, k_2, k_3, k_4}
\delta_{\bf k_1 + k_3 - k_2 - k_4} \nonumber\\ & & \times \psi^{\alpha
\; \dagger}_{ n}({\bf k_1}) \psi^{\alpha}_{ n}({\bf k_2}) \psi^{\beta
\; \dagger}_{ m}({\bf k_3}) \psi^{\beta}_{ m}({\bf k_4}) , 
\end{eqnarray} where
the subscript $n,m$ on the electron operators denote the
Matsubara frequencies $\omega_n, \omega_m$, also {$\sum_{n, m}$ refers to
Matsubara frequencies summation}, and 
$\alpha, \beta = 1, \ldots n$ are replica
indices. 

To derive a low energy effective theory we follow Belitz and
Kirkpatrick~\cite{belitz_1994a} and consider the disorder-induced
interaction with all momenta near the Fermi surface. There are three
possible ways of pairing up the scattering: $(1)$ is the small angle
(or direct) scattering with $\bf k_2 = k_1 + q$, $(2)$ is the large
angle (or exchange) scattering with $\bf k_4 = k_1 + q$, and $(3)$ is
the pair (or $2 k_F$) scattering where $\bf k_3 = k_1 +q$. The
momentum transfer $\bf q$ is now restricted to be small (with a
cut-off much smaller than the Fermi momentum). It can be shown that
type (1) only leads to a renormalization of the chemical potential and
we drop it from now on. Type (3) couples to the superconducting order
parameter, but not to the Pomeranchuk one, and we can  show that
this term does not have any effect on the Pomeranchuk order at mean
field level so we neglect it.  However, for a superconductor, the type
(3) term generates a vertex correction that for an $s$-wave
superconductor cancels the propagator correction due to type (2) term,
thereby rendering it insensitive to disorder (Anderson's
theorem)~\cite{abrikosov_1963a}.  Thus, the disorder-induced interaction is
\begin{eqnarray}
S_{\rm dis} &=& \frac{-1}{4\pi N_0 \tau} \sum_{\alpha,\beta} {T^2}\sum_{n, m}
\sum_{\bf k, p, q}  \nonumber \\
& &\times \psi^{\alpha \; \dagger}_{ n}({\bf k}) \psi^{\alpha}_{ n}({\bf p})
\psi^{\beta \; \dagger}_{ m}({\bf p+q}) \psi^{\beta}_{ m}({\bf k+q}) .
\end{eqnarray}

The full low energy effective action after disorder-averaging is thus
$S = \sum_{\alpha} \left( S_0^{\alpha} + S_{\rm int}^{\alpha} \right) +
S_{\rm dis} $ with
\begin{eqnarray}
S_0^{\alpha} & = & \sum_{\bf k} {T}\sum_{n} \psi^{\alpha \;
\dagger}_{\sigma n}({\bf k}) (-i \omega_n + \epsilon_{\bf k} )
\psi^{\alpha}_{\sigma n}({\bf k}) ,\\ 
S_{\rm int}^{\alpha} & = &
{T^3}\sum_{n_1, n_2, m} \sum_{\bf k, k', q} \frac{ V_q({\bf k}, {\bf k'})
}{2}\\ & &\times \psi^{\alpha \;\dagger}_{n_1+m}({ \bf k+q})
\psi^{\alpha}_{n_1}({ \bf k}) \psi^{\alpha \; \dagger}_{n_2 -m} ({\bf
k'-q}) \psi^{\alpha}_{n_2}({ \bf k'}) .\nonumber
\end{eqnarray}

To decouple the four fermion interaction terms, we introduce the $Q$
matrix via essentially a Hubbard-Stratonovich transformation (generalizing Ref.~\cite{belitz_1997b})
\begin{eqnarray}
\left[Q^{\alpha \beta}_{\genfrac{}{}{0pt}{}{n m} 
{\bf k p}}\right]^j_i = \left[\psi^{\alpha \dagger}_n({\bf k})\right]_i 
\left[\psi^{\beta}_m({\bf p})\right]^j  ,
\end{eqnarray} 
where $i,j$ label the spinor components. 

Since the replica-Q-matrix method has already been comprehensively reviewed in
Ref.~\cite{belitz_1994a},\cite{belitz_1997b}, we here only sketch out its application
to the Pomeranchuk instability in the presence of weak quenched disorder.
Assuming that at the saddle point,
there is replica symmetry and spin symmetry, the homogeneous and
un-retarded ansatz for the saddle-point of this action is
\begin{equation} \label{ansatz}
\left[Q^{\alpha \beta}_{\genfrac{}{}{0pt}{}{n m}{\bf k p}}\right]^i_j = 
\delta_{\alpha \beta} \delta_{n m} \delta_{\rm k p}
\delta_{i j} Q_{n \bf k} ,
\end{equation} 
Note that only one function $Q_{n \bf k}$ is needed here (unlike for magnets
or superconductors), because
both disorder and the quadrupolar interaction  induces a self-energy
$i \Lambda_{n \bf k}$ that enters in the same way into the propagator
renormalization.  $\Lambda_{n \bf k}$ is the Fourier transform dual of $Q_{n \bf k}$ 
~\cite{belitz_1997b} and thus has  the same structure as $Q_{n \bf k}$.

With the ansatz eq.~\ref{ansatz}, the saddle point action becomes
\begin{eqnarray} \label{mfS}
S_{\rm sp} &=& -{\rm Tr} \ln 
\left[ -i \omega_n +\epsilon_{\bf k} + i \Lambda_{n \bf k}\right]
- 2 i {T}\sum_{n \bf k} \Lambda_{n \bf k} Q_{n \bf k} \nonumber\\
 & & + 2 \sum_{\bf k p} {V_0({\bf k, p})} \; {T^2} \sum_{n m} Q_{n \bf k} Q_{m \bf p} \nonumber  \\
 & & + \frac{1}{2\pi N_0 \tau}{T} \sum_{n \bf k p}  
Q_{n \bf k} Q_{n \bf p}  . 
\end{eqnarray}  Note that only the ${\bf q}=0$ component of $V_{\bf q}({\bf k, p})$,
{\emph i.e.} $V_0({\bf k, p}) $
matters, because of the assumption of spatial homogeneity in the saddle point ansatz
$Q_{n \bf k}$. Also we have assumed replica symmetry: we can \emph{ a posteriori}
justify this by noting that as we shall see, there are no indications in the free energy of further instabilities in the Fermi surface distorted phase, unlike in the classical spin glass case
which does demand replica symmetry breaking. Presumably this has to do with the much simpler free energy landscape in the Pomeranchuk case, indicating lack of glassiness in
our system.
 Replica symmetry means we can drop the replica indices from now on.
The saddle point equations $\delta S_{\rm sp}/\delta Q_{n \bf k} = 0 $ and 
$\delta S_{\rm sp}/\delta \Lambda_{n \bf k} = 0 $ give
\begin{eqnarray} 
Q_{n \bf k} &=&  
\frac{1}{i \omega_n -\epsilon_{\bf k} -i \Lambda_{n \bf k}} \; ,
\label{mfQ}
\\
i \Lambda_{n \bf k} &=& \frac{1}{2\pi N_0 \tau} \sum_{\bf p}
 Q_{n \bf p} + \sum_{\bf p} V_0({\bf k p}) {T}\sum_m Q_{m {\bf p}} \; .
\label{mfLambda} 
\end{eqnarray}  

First, lets check that the ansatz eq.~\ref{ansatz} recovers known results. For free electrons
with quenched disorder, setting ${V_0({\bf k, p})}=0$ leads to the standard Born
approximation result; at the saddle point, $Q_{n \bf k}$ is just the
electron propagator with a disorder-induced lifetime $\tau$: $Q_{n \bf k}\approx
G_{n \bf k} = [i \omega_n -\epsilon_{\bf k} +\frac{i}{2\tau}{\rm sgn}(\omega_n)]^{-1} $.
For clean electrons with the quadrupolar interaction, setting $\tau \rightarrow \infty$,
and assuming a spatially homogeneous order parameter
({\it i.e.}, only the ${\bf q} =0$ component of ${V_{\bf q}({\bf k, p})}$
is involved), we define the Pomeranchuk order parameter
\begin{equation} \label{Pomm-gap-def}
\phi_{\bf k} \Delta_0 = 
{T}\sum_{n \bf p} {V_0({\bf k, p})} \left\langle \psi^{\dagger}_{n \bf p}
\psi_{n \bf p} \right\rangle =  
2 \phi_{\bf k} {T}\sum_{n \bf p} g_{} \phi_{\bf p} Q_{n \bf p} .
\end{equation} We then recover the clean case mean field order parameter equation
$\Delta_0 =  2 g_{} \sum_{\bf k} \phi_{\bf k}\; f_T\left(\epsilon_{\bf k} +
\phi_{\bf k}\Delta_0\right)$, where $f_T(x) = [\exp x/T +1]^{-1}$ is the  usual Fermi distribution.

\section{Results}

Now we consider the case of weakly disordered electrons with an interaction
favoring a Pomeranchuk instability.
eqs.~\ref{Pomm-gap-def},\ref{mfQ},\ref{mfLambda} lead to the order parameter equation
\begin{eqnarray} \label{dirtyDelta0}
\Delta_0 = 
%2 g_{} \sum_{n \bf k} \phi_{\bf k} G_{n \bf k} = 
2 g_{} T\sum_{n \bf k} \phi_{\bf k}
\frac{1}{i \omega_n - \epsilon_{\bf k} + i \frac{1}{2\tau} 
{\rm sgn} \omega_n - 
\phi_{\bf k}\Delta_0} .
\end{eqnarray} 
This is in fact the non $s$-wave, non-magnetic analogue of the Stoner
instability of a ferromagnet.  The extra
angular dependence in the momentum sum means that this order parameter
equation does not reduce to the clean case.
(By contrast, the $s$-wave Stoner Pomeranchuk instability
is unaffected, to leading order, by
impurities~\cite{kirkpatrick_2000a}). Thus, just as for non $s$-wave
superconductors, there is no Anderson's theorem for $l\neq 0$
Pomeranchuk instabilities, because of its angular dependence in
momentum space. eq.~\ref{dirtyDelta0} simplifies to:
%using the usual digamma function identity 
%$\sum_{n=0}^{N-1} \frac{1}{n + x} = \psi(x+N) - \psi(x)$ 
\begin{eqnarray}
\Delta_0 = \frac{2 g_{}}{\pi} \sum_{\bf k} \phi_{\bf k} \; {\rm Im}
\psi\left(\frac{1}{2} + \frac{1}{4 \pi T \tau} - i \frac{\epsilon_{\bf
k} + \Delta_0 \phi_{\bf k}}{2 \pi T}\right) . 
\label{dirtyDelta}
\end{eqnarray} 
As expected, disorder smears out the Fermi distribution to give the
digamma function $\psi(x)$: crudely speaking, disorder raises the
effective temperature.

In contrast to weak-coupling superconductivity where there is always a
second-order transition for arbitrary weak interactions, the Pomeranchuk
instability requires a critical coupling and the transition can either
be first or second-order. To check the order of the transition, we need to evaluate the free energy. Substituting the mean
field equations \ref{mfQ},\ref{mfLambda} into the saddle point action
eq.~\ref{mfS}, together with the approximation (just as for disordered
free electrons case) $\sum_{\bf p} Q_{n {\bf p}} \approx -i \pi N_0
{\rm sgn} \omega_n$ and the definition of the order parameter parameter 
(eq. ~\ref{Pomm-gap-def})
%$\Delta_0 = 2 g_{} \sum_{\bf k} \phi_{\bf k} \sum_n Q_{n {\bf k}}$
, the saddle point free energy becomes
\begin{eqnarray}
F_{\rm sp} = - {\rm Tr} \ln 
\left[ -i\left(\omega_n + \frac{{\rm sgn} \omega_n}{2 \tau}\right) +
{\tilde \epsilon}_{\bf k} -\mu \right] 
-\frac{\Delta_0^2}{2 g_{}} + {\rm cst}  ,\nonumber
%\label{Fsp}
\end{eqnarray} 
where the renormalized dispersion ${\tilde \epsilon}_{\bf k} =
\epsilon_{\bf k} +\mu +\Delta_0 \phi_{\bf k}$.  Defining the
renormalized density of state $N_{\Delta}(\epsilon) = \sum_{\bf k}
\delta(\epsilon -{\tilde \epsilon}_{\bf k})$, 
%to give
%\begin{eqnarray} \label{Fsp-DOS}
%F_{\rm sp} &= & 4 T \int_{-4 t}^{4 t} d \epsilon \; N_{\Delta}(\epsilon)
%{\rm Re} \ln  \Gamma \left( \frac{1}{2} + \frac{1}{4 \pi T \tau} +
% i \frac{\epsilon-\mu}{2 \pi T} \right) \nonumber\\
%& & -\frac{\Delta_0^2}{2 g_{}} + {\rm cst}  ,
%\end{eqnarray} 
the order parameter equation \ref{dirtyDelta0} becomes
\begin{eqnarray} \label{gap-DOS}
\frac{\Delta_0}{g_{}} = 4 T \int_{-4 t}^{4 t} d \epsilon \; 
\frac{\partial N_{\Delta}(\epsilon)}{\partial\Delta_0}
{\rm Re} \ln \Gamma \left( \frac{1}{2} + 
\frac{1}{4 \pi T \tau} +i \frac{\epsilon-\mu}{2 \pi T} \right).
\end{eqnarray}

For  quantitative results for effects of disorder,
we evaluate the free energy and mean field equations numerically for
the 2D square lattice with bare dispersion $\epsilon_{\bf k} = -2 t
\left( \cos k_x + \cos k_y \right) - \mu $, and the d-wave form
factor: $\phi_{\bf k} = \cos k_x - \cos k_y$, with the corresponding
renormalized density of states (see Ref.~\cite{kee_2003b,khavkine_2004a}
for the actual form).
%\begin{eqnarray} 
%\label{D-DOS}
%N_{\Delta}(\epsilon) = \frac{1}{2 t \pi^2} {\rm Re} 
%\frac{1}{\sqrt{1-(\epsilon/4t)^2}} 
%{\bf K}\left( \frac{1-(\Delta_0 /2t)^2}{1-(\epsilon/4t)^2} \right), 
%\end{eqnarray} 
%and the density of states is zero outside the bandwidth of $-4t$ to
%$4t$. We use the notation of Ref.~\cite{abramowitz_1964a} for the
%elliptic integral of the first kind ${\bf K}$. 
In the  following, energies and scattering rates, $\tau^{-1}$, 
are measured in units of $2t$.

\begin{figure}[tb]
%\begin{center}
\includegraphics[width=8.5cm]{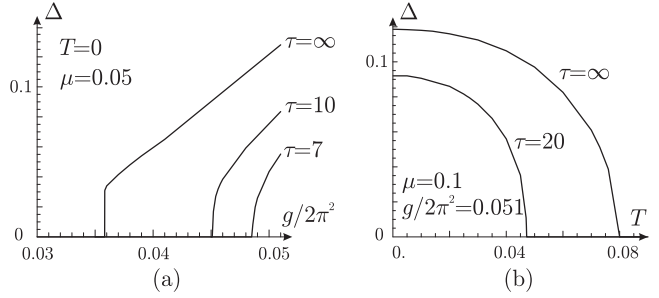}
\caption{The effect of disorder on the magnitude of the order
  parameter $\Delta$ (all energies and scattering rates expressed in units of
  $2t$). (a) At $T=0$ as disorder decreases the lifetime, $\tau$, the critical
  coupling, $g/2\pi^2$ increases and a first order transition becomes
  second order. (b) At fixed coupling, the order parameter and
 $T_c$ are rapidly suppressed   by weak disorder.} 
\label{orderparm}
\end{figure}
First we consider how disorder changes the evolution of the order
parameter. In Fig.~\ref{orderparm}(a) we see  for fixed chemical
potential, $\mu$, how the Pomeranchuk order parameter is modified at
$T=0$ by scattering. Disorder both increases the critical
coupling and turns at $T=0$ from first order in the clean limit, to
second order. In Fig.~\ref{orderparm}(b) we see that rather weak
disorder dramatically reduces $T_c$. 

\begin{figure}[tb]
\onefigure[width=8.5cm]{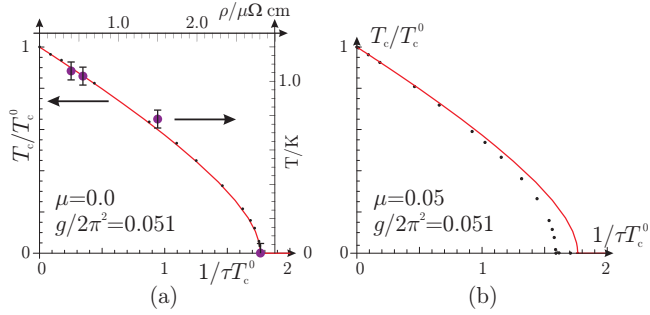}
\caption{The Pomeranchuk transition temperature $T_c$ relative
 to the clean value $T_c^0$, as a function of dimensionless disorder
 $1/\tau T_c^0$ at a parameter regime  where the Pomeranchuk transition is second order. (a) At $\mu=0$ the
 numerical solution of the order parameter equation ~\ref{gap-DOS} 
 (small points) coincides with the Abrikosov-Gorkov result of eq.~\ref{AG-Tc}
 (solid line). The large data points show a fit to published data~\cite{perry_2004b, grigera_2004b} on ${\rm Sr_3Ru_2O_7}$ at 7.95T where the resistivity minimum and the 
zero-field residual resistivity indicated $T_c$ and $1/\tau$ respectively.
(b) At $\mu=0.05$ there are small
 deviations from the Abrikosov-Gorkov form. 
 These are corrections from the
 weak-coupling approximation and are related to the detailed
 structure of the density of states we used.
 }
\label{tcdis}
\end{figure}

We next consider the disorder dependence of the Pomeranchuk
transition temperature, $T_c$. In Fig.~\ref{tcdis}(a) we
choose a parameter region where the clean transition is second order 
($g_{}/2t2\pi^2=0.051$ and $\mu/2t =0, 0.05$). 
Since the order parameter goes smoothly to zero, we can
simplify the order parameter equation (eq.~\ref{dirtyDelta0}) 
%by noting that the right
%hand side of eq.~\ref{dirtyDelta0} must be real, so we can take directly the real part 
to get the second order transition order parameter equation
\begin{eqnarray} \label{gap-2nd}
-\frac{1}{g_{}} = {T}\sum_{n,{\bf k}} \phi_{\bf k}^2 \frac{1}{(\omega_n +
 \frac{1}{2\tau} {\rm sgn} \omega_n)^2 + \epsilon_{\bf k}^2} .
\end{eqnarray} 
We then note that this order parameter equation is {\it identical} to  the one
determining the critical temperature for $d$-wave superconductor with
non-magnetic disorder. Note that this is true only for the $T_c$
equation for a second order transition: there are first order
transitions at larger $\mu$, and furthermore, the full order parameter equation
for the disordered Pomeranchuk instability has a different form to
the d-wave superconductor with disorder.

Thus, from eq.~\ref{gap-2nd}, we get the familiar Abrikosov-Gor'kov 
form~\cite{abrikosov_1961a} for the disordered gap equation 
\begin{eqnarray} \label{AG-Tc}
\ln \left(\frac{T_{c0}}{T_c}\right) = 
\psi\left(\frac{1}{2} + \frac{1}{4 \pi T_c \tau}\right)
 - \psi\left(\frac{1}{2}\right) ,
\end{eqnarray} 
In figs.~\ref{tcdis} (a) and (b), the solid curve is this universal
Abrikosov-Gorkov form, while the small data points are direct numerical
evaluation of the {\it general} (i.e. not just for second order
transition) order parameter equation \ref{gap-DOS}. 
At $\mu/2t = 0$, the direct evaluation
coincide with the universal form, while for finite chemical potential
$\mu/2t = 0.05$ shown in Fig.~\ref{tcdis} (b), some small deviation
can be seen at larger disorder, due to the
approximation in the radial ${\bf k}$-integral that goes into deriving
eq.~\ref{AG-Tc}.

We have also taken existing data  of  on ${\rm Sr_3Ru_2O_7}$ at 7.95T 
\cite{perry_2004b, grigera_2004b} where the resistivity minimum and the  zero-field 
residual resistivity are taken to indicate $T_c$ and $1/\tau$ respectively, and plotted 
them as large dots in Fig.~\ref{tcdis}(a). The reasonable fit shows that the putative transition in ${\rm Sr_3Ru_2O_7}$ (which experimentally, is found not to be of superconductivity type)
does follow the universal Abrikosov-Gorkov form for disordered Pomeranchuk transition,
even if our actual model interaction of eq. 1 may be too simplistic.
\begin{figure}[tb]
\onefigure[width=8.5cm]{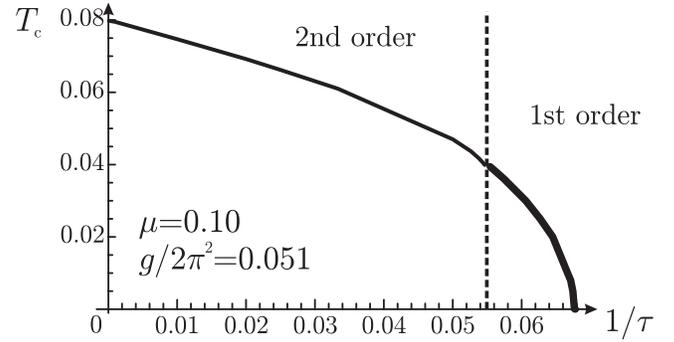}
\caption{Suppression of $T_c$ as a function of scattering rate 
in units of $2t$ for a Fermi surface away from the van Hove point.
The transition is driven first order at low temperatures.}
\label{1storderTc}
\end{figure}

Finally we consider the disorder dependence on the transition
temperature where, at low temperatures, the transition can become
first order~\cite{kee_2003a} such as when the system is further away
from half-filling, {\it e.g.} with $\mu= 0.10$ (Fig.~\ref{1storderTc}).
Then, with larger disorder (larger $1/\tau$), the transition turns
from second to first order (bold line). Surprisingly, the effect of
increasing disorder is opposite to increasing $T$.  Higher $T$ smears
out the Fermi function and leads to a smaller order parameter and eventually a
second order transition results.\cite{khavkine_2004a} Of course, it is  possible
that including Gaussian fluctuations around the saddle point may turn first order
transitions into second order ones in the presence of disorder. Future work is needed
to resolve this issue. What is clear is that even at the mean field level, there is
a strong suppression of $T_c$ for the Pomeranchuk instability with increasing
disorder. 

In summary, we have calculated the strong  dependence of the
$d$-wave Fermi surface distortion transition on dilute disorder and shown that the effect
is reminiscent both qualitatively {\em and} quantitatively of the
strong dependence on impurities of the order parameter and $T_c$ in non $s$-wave
superconductors. 
%Although we have here used the replica mean field
%method, we have also obtained the same results from a diagrammatic approach. 
Our results suggest that detailed disorder dependence of
``hidden order'' transitions could be used to indicate Pomeranchuk
type order just as is the case for low $T_c$ unconventional
superconductors. The extreme sensitivity to
disorder~\cite{grigera_2004b} of the low-temperature metal-to-metal
transition in ${\rm Sr_3Ru_2O_7}$ is suggestive of this 
[see Fig.~\ref{tcdis}(a)].
Our theoretical results are intended to motivate
a more comprehensive systematic experimental study to compare with our 
quantitative predictions for the putative ordered state in ${\rm Sr_3Ru_2O_7}$.

\acknowledgments
We are grateful to the EPSRC and the Leverhulme Trust for their
financial support, and to Eduardo Fradkin and Derek Lee for useful discussion. 

%\bibliographystyle{eplbib}
%\bibliography{pomdis-epl}

\end{document}